\documentclass{article}
\usepackage{amsfonts,amsmath,tikz,multirow,bm,overpic} 
\usepackage[breaklinks]{hyperref}
\usepackage[accepted]{icml2013}
\usetikzlibrary{arrows,positioning} 
\icmltitlerunning{Gaussian Process Vine Copulas for Multivariate Dependence}

\newcommand{\f}[1]{\bm{#1}}

\begin{document} 
\twocolumn[
  \icmltitle{Gaussian Process Vine Copulas for Multivariate Dependence}
  
  \icmlauthor{David Lopez-Paz}{david.lopez@tue.mpg.de}
  \icmladdress{Max Planck Institute for Intelligent Systems}
  \icmlauthor{Jose Miguel Hern\'andez-Lobato}{jmh233@eng.cam.ac.uk}
  \icmlauthor{Zoubin Ghahramani}{zoubin@eng.cam.ac.uk}
  \icmladdress{University of Cambridge}
  
  \icmlkeywords{copulas, 
                expectation propagation,
                vines,
                conditional dependence,
                density estimation}
  
  \vskip 0.3in
]

\begin{abstract} 
\emph{Copulas} allow to learn marginal distributions separately from the
multivariate dependence structure (copula) that links them together into a
density function.  \emph{Vine factorizations} ease the learning of
high-dimensional copulas by constructing a hierarchy of conditional bivariate
copulas. However, to simplify inference, it is common to assume that each of
these conditional bivariate copulas is independent from its conditioning
variables.  In this paper, we relax this assumption by discovering the latent
functions that specify the shape of a conditional copula given its conditioning
variables.  We learn these functions by following a Bayesian approach based on
sparse Gaussian processes with expectation propagation for scalable,
approximate inference. Experiments on real-world datasets show that, when
modeling all conditional dependencies, we obtain better estimates of the
underlying copula of the data.
\end{abstract} 

\section{Introduction}
Copulas are becoming a popular approach in machine learning to describe
multivariate data \cite{Elidan2012,Kirshner2007,Elidan2010,Wilson2010}.
Estimating multivariate densities is difficult due to possibly complicated forms
of the data distribution and the curse of dimensionality. Copulas simplify this
process by separating the learning of the marginal distributions from the
learning of the multivariate dependence structure, or \emph{copula}, that links
them together into a density model \cite{Joe2005}. Learning the
marginals is easy and can be done using standard univariate methods. However,
learning the copula is more difficult and requires models that can represent a
broad range of dependence patterns. For the two-dimensional case, there exists
a large collection of parametric copula models \cite{Nelsen2006}. However, in
higher dimensions, the number and expressiveness of families of parametric
copulas is more limited. A solution to this problem is given by pair copula
constructions, vine copulas or simply \emph{vines}
\cite{Bedford2002,Kurowicka2006}.  These are graphical models that decompose any
multivariate copula into a hierarchy of bivariate copulas, where some of them
will be conditioned on a subset of the data variables.  The deeper a bivariate
copula is in the vine hierarchy, the more variables it will be conditioned on.
If the conditional dependencies described above are ignored, vines are a
straightforward approach to construct flexible high-dimensional dependence
models using standard parametric bivariate copulas as building blocks.

\begin{figure}
  \begin{center}
    \includegraphics[width=\linewidth]{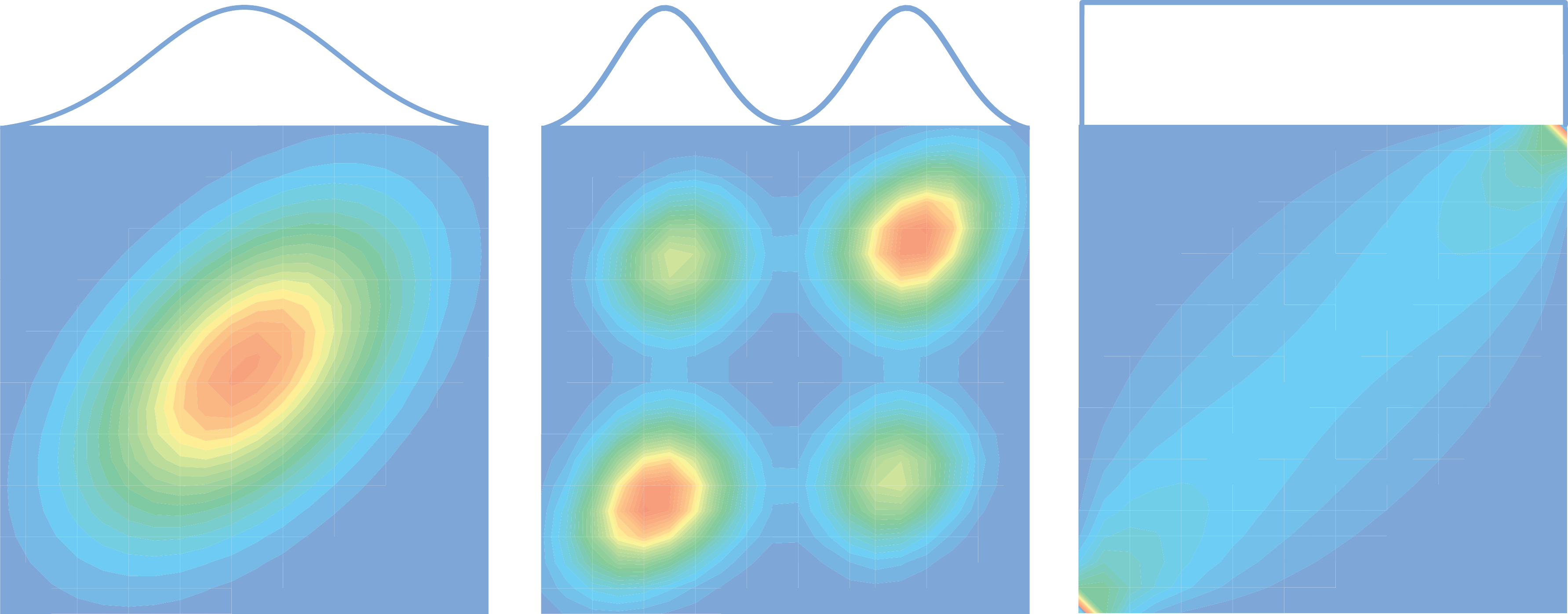}
    \label{fig:models}
    \caption{Two bidimensional densities (left, middle) that share the same
    underlying Gaussian copula, with correlation parameter $\theta = 0.8$
    (right).  The two distributions differ because of their
    marginal distributions, depicted at the top of each density plot.}
  \end{center}
\end{figure}

The impact of ignoring conditional dependencies in the copula functions is
likely to be problem specific.  \citet{Hobaek2010} show thorough experiments
with synthetic data that, in specific cases, ignoring conditional dependencies
can lead to reasonably accurate approximations of the true copula. By contrast,
Acar, Genest and Neslehova \yrcite{Acar2012} indicate that this
\emph{simplifying assumption} can be in other cases misleading,
and develop a method to condition parametric bivariate copulas on a single
scalar variable. In this paper, we extend the work of \citet{Acar2012} and
propose a general technique to construct arbitrary vine models with full
conditional parametric bivariate copulas. Our results on several real-world
datasets show that it is often important to take into account 
conditional dependencies when constructing a vine model.

The proposed method is based on the fact that most parametric bivariate copulas
can be specified in terms of Kendall's rank correlation parameter $\tau \in
[-1,1]$ \cite{Joe1997}. The dependence of the copula on a vector of
conditioning variables $\mathbf{u}=(u_1,\ldots,u_d)^\text{T}$ is then captured
by specifying the relationship $\tau = \sigma(f(\mathbf{u}))$, where  $f :
\mathbb{R}^d \rightarrow \mathbb{R}$ is a non-linear function and $\sigma :
\mathbb{R} \rightarrow [-1,1]$ is a scaling operation.  We follow a Bayesian
approach to learn $f$ from available data.  In particular, we place a Gaussian
process (GP) prior on $f$ and use expectation propagation for approximate
inference \cite{Rasmussen2006,Minka2001}. To make our method scalable, we use
sparse GPs based on the generalized FITC approximation
\cite{Snelson2006,Naish-Guzman2007}.  

\section{Copulas and Vines}

When the components of a $d$-dimensional random vector
$\mathbf{x}=(x_1,\ldots,x_d)^\text{T}$ are independent, their density function
$p(\mathbf{x})$ can be factorized as
\begin{equation}
p(\mathbf{x}) = \prod_{i=1}^{d} p(x_i)\,.\label{eq:indep}
\end{equation}
The previous equality does not hold when $x_1,\ldots,x_d$ are not independent.
Nevertheless, the differences can be corrected by multiplying the right hand
side of (\ref{eq:indep}) by a specific function that fully describes any
possible form of dependence between the random variables $x_1,\ldots,x_d$. This
function is called the \emph{copula} of $p(\mathbf{x})$ \cite{Nelsen2006}, and
satisfies:
\begin{equation}
p(\mathbf{x}) =
\left[\prod_{i=1}^{d} p(x_i)\right]
\underbrace{c(P(x_1),..., P(x_d))}_\text{copula},
\label{eq:copulaDensity}
\end{equation}
where $P(x_i)$ is the marginal cumulative distribution function (cdf) of the
random variable $x_i$.  The copula $c$ is the joint multivariate density of
$P(x_1),\ldots,P(x_d)$ and it has uniform marginal distributions, since $P(x)
\sim \mathcal{U}[0,1]$ for any random variable $x$ \cite{CaseBerg2001}.  This
non-linear transformation from $x$ to $P(x)$ is known as the
Probability Integral Transform (PIT).  The copula is the density of
$\mathbf{x}$ after eliminating all the marginal information by applying the PIT
to each individual component of $\mathbf{x}$.  Therefore, $c$ describes any
dependence patterns which do not depend on the marginal distributions. If every
$P(x_i)$ is continuous, then $c$ is unique for any $p(\mathbf{x})$
\cite{sklar}. However, infinitely many multivariate distributions share the same
underlying copula (Figure \ref{fig:models}). 

The main advantage of copulas is that they separate the learning of univariate
marginal distributions from the learning of the multivariate dependence
structure that describes how they are coupled \cite{Joe2005}.  Learning the
marginals is easy and can be done using standard univariate methods. However,
learning the copula is more difficult and requires models that can represent a
broad range of dependence patterns.  For the two-dimensional case, a large
collection of parametric copula models is available \cite{Nelsen2006}.  Some
examples are the Gaussian, Student, Clayton, Independent, Gumbel or Frank
copulas.  Each of these families describes a different dependence structure
between two random variables. An intuitive example is the copula that describes
independence, that is, the independent copula: it has density constant and
equal to one, as one can infer from equations (\ref{eq:indep}) and
(\ref{eq:copulaDensity}).  The Appendix contains more on the bivariate Gaussian
Copula, which is used extensively used throughout this paper. 

Although there exist many parametric models for two-dimensional copulas, for more
than two dimensions the number and expressiveness of families of parametric
copulas is more limited. A solution to this problem is given by pair copula
constructions, vine copulas or simply \emph{vines}
\cite{Joe1996,Bedford2002,Kurowicka2006}.

\subsection{Regular Vines}\label{sec:vines}

\emph{Vine copulas} are hierarchical graphical models that 
factorize a $d$-dimensional copula density into a product of $d(d-1)/2$
bivariate conditional copula densities. They offer great modeling flexibility,
since each of the bivariate copulas in the factorization can belong to a
different parametric family. Several types of vines have been proposed in the
literature. Some examples are canonical vines (C-Vines), drawable vines
(D-vines) or regular vines (R-Vines). In this paper we focus on regular vines,
since they are a generalization of all the other types \cite{rvines}.

An R-vine $\mathcal{V}$ specifies a factorization of a copula density
$c(u_1,\ldots,u_d)$ into a product of bivariate conditional copulas. Such
R-vine is constructed by forming a nested set of $d-1$ undirected trees, in
which each of their edges corresponds to a conditional bivariate copula density.
A particular nested set of trees identifies a
particular valid factorization of $c(u_1,\ldots,u_d)$. These trees can be
sequentially constructed as follows:

\begin{enumerate}
\item Let $T_1, \ldots, T_{d-1}$ be the trees in a R-Vine $\mathcal{V}$, each
of them with set of nodes $V_i$ and set of edges $E_i$. 
\item Every edge $e\in E_i$ has associated three sets of variable indexes
$C(e),D(e),N(e)\subset \{ 1, \ldots, d \}$ called the conditioned, conditioning
and constraint sets of $e$, respectively.
\item The first tree in the hierarchy has set of nodes $V_1 = \{1,\ldots,d\}$ and
set of edges $E_1$, which is obtained by inferring a spanning tree from a
complete graph $G_1$ over $V_1$.
\item For any edge $e \in E_1$ joining nodes $j,k\in V_1$, $C(e) = N(e) = \{j,k\} $
and $D(e) = \{ \emptyset \}$. 
\item The $i$-th tree has node set $V_i = E_{i-1}$ and edge set $E_i$, for $i = 
2,\ldots d-1$. $E_i$ is obtained by inferring a spanning tree from a
graph $G_i$; this graph has set of nodes $V_i$ and edges $e=(e_1,e_2)\in E_i$,
such that $e_1, e_2 \in E_{i-1}$ share a common node in $V_{i-1}$.
\item Edges $e=(e_1,e_2)\in E_i$ have conditioned,
conditioning and constraint sets given by $C(e) = N(e_1) \Delta N(e_2)$, $D(e)=
N(e_1) \cap N(e_1)$ and $N(e) = N(e_1)\cup N(e_2)$, where
$A\,\Delta\,B = (A \setminus B) \cup (B \setminus A)$.
\end{enumerate}

\begin{figure}
  \begin{center}
\resizebox{\linewidth}{!}
{
  \tikzset{
    select/.style={draw=black,line width=1.2mm},
    punkt/.style={circle,draw=black,minimum height=4em,text centered}
  }
  \begin{tikzpicture}
    \node[punkt]             (1) {$1$};
    \node[punkt, right=of 1] (2) {$2$};
    \node[punkt, below=of 1] (3) {$3$};
    \node[punkt, below=of 2] (4) {$4$};
    \path[select] (1) edge node[rotate=90,anchor=south,shift={(0mm,5mm)}]{$e_1 = 1,3|\emptyset$} (3);
    \path[select] (2) edge node[rotate=45,anchor=north, shift={(-0mm,0mm)}]{$e_2 = \,\,2,3|\emptyset$} (3);
    \path (1) edge (4);
    \path (1) edge (2);
    \path (2) edge node[anchor=west]{} (4);
    \path[select] (3) edge node[anchor=north, shift={(0,-5mm)}]{$e_3 = 3,4|\emptyset$} (4);
    \node[punkt, right=0.5 of 2] (13) {$1,3|\emptyset$};
    \node[punkt, right=of 13] (23) {$2,3|\emptyset$};
    \node[punkt, below= of 13] (34) {$3,4|\emptyset$};
    \path[select] (13) edge node[anchor=south,above, shift={(0,5mm)}]{$e_4 = 1,2|3$} (23);
    \path[select] (13) edge node[rotate=90,anchor=south,shift={(0,5mm)}]{$e_5 = 1,4|3$} (34);
    \path[] (34) edge node[anchor=east]{} (23);
    \node[punkt, right=0.5 of 34] (231) {$1,2|3$};
    \node[punkt, right=of 231,] (341) {$1,4|3$};
    \path[select] (231) edge node[anchor=south,above, shift={(0,5mm)}]{$e_6 = 2,4|1,3$} (341);
    \node[above=0.1 of 1, shift={(12mm,0)}]{\underline{$G_1/\bm{T_1}$}};
    \node[above=0.3 of 13, shift={(12mm,0)}]{\underline{$G_2/\bm{T_2}$}};
    \node[above=0.3 of 231, shift={(12mm,0)}]{\underline{$G_3/\bm{T_3}$}};
  \end{tikzpicture}
}
\vskip -0.6 cm
\begin{equation*}
    c_{1234} = 
    \underbrace{
    \underbrace{c_{13|\emptyset}}_{e_1}
    \underbrace{c_{23|\emptyset}}_{e_2}
    \underbrace{c_{34|\emptyset}}_{e_3}}_{T_1}
    \underbrace{
    \underbrace{c_{12|3}}_{e_4}
    \underbrace{c_{14|3}}_{e_5}}_{T_2}
    \underbrace{
      \underbrace{c_{24|13}}_{e_6}}_{T_3}
\end{equation*}
  \end{center}
\vskip -0.3 cm
\caption{Example of the hierarchical construction of an R-vine factorization of
a copula density of four variables $c(u_1,u_2,u_3,u_4)$. The edges selected to
form each tree are highlighted in bold. Conditioned and conditioning sets for
each node and edge are shown as $C(e) | D(e)$.}
\label{fig:vine_construction}
\end{figure}
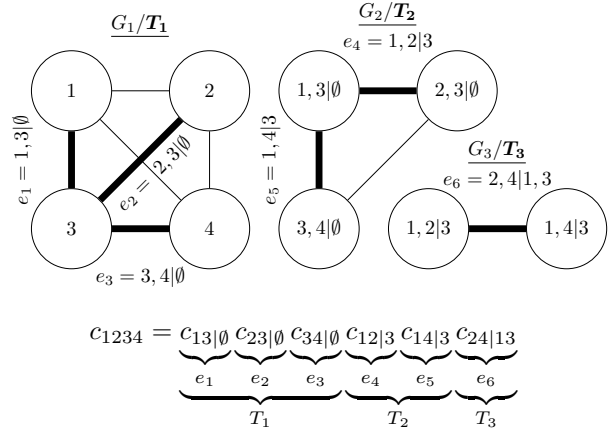

Each of the edges in the trees $T_1,\ldots,T_{d-1}$ forming the vine
$\mathcal{V}$ is a different factor in the factorization of
$c(u_1,\ldots,u_d)$, i.e. a different conditional bivariate copula density.
Since there are a total of $d(d-1)/2$ edges, $\mathcal{V}$ factorizes
$c(u_1,\ldots,u_d)$ as the product of $d(d-1)/2$ factors. We now show how to
obtain the form of each of these factors.  For any edge $e(j,k) \in T_i$ with
conditioned set $C(e)=\{j,k\}$ and conditioning set $D(e)$ we define
$c_{jk|D(e)}$ to be the bivariate copula density for $u_j$ and $u_k$ given the
value of the conditioning variables $\{u_i: i \in D(e) \}$, that is,
\begin{align} c_{jk|D(e)} & := c( P_{j|D(e)}, P_{k|D(e)}| u_i : i \in
D(e)),\label{eq:defcjk} \end{align} where $P_{j|D(e)}:=P(u_j|u_i : i \in D(e))$
is the conditional cdf of $u_j$ given the value of the conditioning variables
$\{u_i: i \in D(e) \}$.  Then, the vine $\mathcal{V}$ formed by the hierarchy
of trees $T_1,\ldots,T_{d-1}$ specifies the following factorization for the
copula density:
\begin{equation}
c(u_1, \ldots, u_d) = \prod_{i=1}^{d-1} \prod_{e(j,k) \in E_i}
c_{jk|D(e)}\,,\label{eq:vindeDecomposition}
\end{equation}
as shown by \citet{Kurowicka2006}. 

Figure \ref{fig:vine_construction} exemplifies how to construct a regular vine that
factorizes the copula density $c(u_1, u_2, u_3, u_4)$ into the product of 6
bivariate conditional copula densities.  The first tree $T_1$ has node set $V_1
= \{ 1,2,3,4 \}$.  The edge set $E_1$ is obtained by selecting a spanning tree
over $G_1$, the complete graph for the nodes in $V_1$.  Our choice for $E_1$ is
highlighted in bold in the left-most plot in Figure
\ref{fig:vine_construction}.  Edges in $E_1 = \{ e_1, e_2, e_3
\}$ have conditioned and constraint sets $C(e_1) = N(e_1) = \{ 1, 3 \}$,
$C(e_2) = N(e_2) = \{2,3\}$, $C(e_3) = N(e_3) = \{3,4\}$ and conditioning sets
$D(e_1) = D(e_2) = D(e_3) = \{ \emptyset \}$.  The second tree in the hierarchy
has node set $V_2 = E_1$.  In this case, we select a spanning tree over $G_2$,
a graph with node set $V_2$ and edge set formed by pairs of edges $e_i, e_j \in
E_1$ sharing some common node $v_k \in V_1$.  We select $E_2 = \{ e_4, e_5 \}$
with conditioned sets $C(e_4) = N(e_1) \Delta N(e_2) = \{1,2\}$ and $C(e_5) =
N(e_2) \Delta N(e_3) = \{1,4\}$, conditioning sets $D(e_4) = N(e_1) \cap N(e_2)
= \{ 3 \}$ and $D(e_5) = N(e_2) \cap N(e_3) = \{3\}$, and constraint sets
$D(e_4) = N(e_1) \cup N(e_2) = \{ 1, 2, 3 \}$ and $D(e_5) = N(e_1) \cap N(e_3)
= \{ 1, 3, 4 \}$.  Finally, we build a third graph $G_3$ with node set $V_3 =
E_2$ and only one edge $e_6$. This last edge is the only possible spanning tree
and has node set $V_3$ and edge set $E_3=\{e_6\}$.  The edge $e_6$ has
conditioned set $C(e_6) = N(e_4) \Delta N(e_5) = \{2,4\}$, conditioning set
$D(e_6) = N(e_4) \cap N(e_5) = \{1,3\}$ and constraint set $N(e_6) = N(e_4)
\cup N(e_5) = \{ 1,2,3,4\}$.  The resulting factorization of
$c(u_1,u_2,u_3,u_4)$ given by the tree hierarchy is shown at the bottom of
Figure \ref{fig:vine_construction}.

There exist many factorizations of a copula density $c(u_1,\ldots,u_d)$ in
terms of bivariate copulas.  Each factorization is determined by the specific
choices of the spanning trees $T_1,\ldots,T_d$ in the algorithm described
above.  In practice, the trees are selected by assigning a weight to each edge
$e(j,k)$ (copula $c_{jk|D(e)}$) in the graphs $G_1,\ldots,G_{d-1}$ and then
selecting the maximum spanning tree at each iteration.  A common practice is to
directly relate the weight of the edge $e(j,k)$ to the amount of dependence
described by the corresponding copula $c_{jk|D(e)}$. This amount of dependence
can be measured as the absolute value of the empirical Kendall's $\tau$
correlation coefficient between the samples of $u_{j|D(e)}$ and
$u_{k|D(e)}$.  The maximum spanning tree can then be selected efficiently using
Prim's algorithm \cite{prim,rvines}.

On the first tree of a vine, only pairwise dependencies are described, and the
corresponding copulas are not conditioned. The following trees describe
dependencies between 3, 4, ... and $d-1$ variables by means of increasingly deeper
conditioning, until completing a full description of the
joint $d-$dimensional copula density.  Since the cost of constructing the full
tree hierarchy is quadratic in $d$, one may choose to prune the vine and
construct only the first $d' < (d-1)$ trees, ignoring the remaining copula
densities in the factorization. Since the independent copula has pdf constant
and equal to one, this pruning assumes independence in the higher order
interactions captured by the ignored copulas.

\subsection{Conditional Dependencies in Vines}

As shown in equations (\ref{eq:defcjk}) and (\ref{eq:vindeDecomposition}), vine
distributions require to calculate marginal conditional cdfs and conditional
bivariate copula densities.  The number of variables to 
condition on increases as we move deeper in the vine hierarchy. In general, to
obtain the factors corresponding to the $i$-th tree, we have to condition both
copula densities and marginal cdfs to $i-1$ variables. The computation of the
conditional cdfs appearing at tree $T_{i}$ can be done using the copula
functions from the previous tree $T_{i-1}$.  In particular, the
following recursive relationship holds
\begin{equation}
  P_{j|D(e)} = \frac{\partial C_{{jk}|D(e)\setminus \{k\}}}{\partial
  P_{k|D(e)\setminus \{k\}}},\label{eq:h_function}
\end{equation}
where $C_{{jk}|D(e)\setminus \{k\}}:=C(P_{j|D(e)\setminus
\{k\}},P_{k|D(e)\setminus k}|u_i: i \in D(e)\setminus \{k\})$ is the cdf of the
conditional copula density $c_{{jk}|D(e)\setminus \{k\}}$ and $D(e)\setminus
\{k\}$ denotes the conditioning set $D(e)$ with the element $k$ removed
\cite{Joe1996}. This derivative has well-known, closed-forms for each
parametric copula (refer to the Appendix for the Gaussian copula case).
However, we still have to compute the conditional bivariate copula densities.
A solution commonly found in the literature is to assume that the copulas
$c_{jk|D(e)}$ in (\ref{eq:vindeDecomposition}) are independent of their
conditioning variables. This is known as the \emph{simplifying assumption} for
vine copulas \cite{Hobaek2010}.  The main advantage is that we can construct
vine models using standard unconditional parametric copulas.  The disadvantage
is that we may fail to capture some of the dependencies present in the data.
As an alternative to the simplifying assumption, we now present a general
technique to construct conditional parametric bivariate copulas.

\section{Proposed Approach to Estimate Conditional Bivariate Copulas}\label{sec:ccop}

In this section we address the estimation of the conditional copula of two
random variables $X$ and $Y$ given a vector of conditioning variables
$\mathbf{Z}=(Z_1,\ldots,Z_d)^\text{T}\in \mathbb{R}^d$.  Let $P_{X|\mathbf{Z}}$
and $P_{Y|\mathbf{Z}}$ be the conditional cdfs of $X$ and $Y$ given
$\mathbf{Z}$.  \citet{Patton2006} shows that the conditional copula of $X$ and
$Y$ given $\mathbf{Z}$ is the conditional distribution of the random variables
$U=P_{X|\mathbf{Z}}(X|\mathbf{Z})$ and $V=P_{Y|\mathbf{Z}}(Y|\mathbf{Z})$ given
$\mathbf{Z}$.  We assume a parametric bivariate copula for the joint
distribution of $U$ and $V$.  This type of copulas can often be fully specified
in terms of Kendall's $\tau$ rank correlation coefficient \cite{Joe1997}.
Table \ref{table:thetatau} shows, for some widely-used copula families, the domain of their
parameter $\theta$ and the corresponding bijective expressions for $\theta$ as
a function of Kendall's $\tau$. To capture the dependence of the copula on
$\mathbf{Z}$ we introduce a latent function $g:\mathbb{R}^d\rightarrow[-1,1]$
such that $\tau = g(\mathbf{Z})$.  The task of interest is then to estimate $g$
given observations of $X$, $Y$ and $\mathbf{Z}$.

When $P_{X|\mathbf{Z}}$ and $P_{Y|\mathbf{Z}}$ are known, we can transform any
sample of $X$, $Y$ and $\mathbf{Z}$ into a corresponding sample of $U$, $V$ and
$\mathbf{Z}$.  Let $\mathcal{D} = \{ \mathcal{D}_{U,V} = \{(u_i,
v_i)\}_{i=1}^n, \mathcal{D}_\mathbf{Z} = \{\mathbf{z}_i\}_{i=1}^n\}$ be such a
sample, where $u_i$ and $v_i$ and $\mathbf{z}_i$ are paired.  To guarantee that
$g(x) \in [-1,1]$, we assume w.l.o.g. that $g(x) = 2\Phi(f(\textbf{x})) - 1$, where
$\Phi(x)$ is the standard Gaussian cdf and $f:\mathbb{R}^d\rightarrow
\mathbb{R}$ is a non-linear function that uniquely specifies $g$. We can
infer $g$ by placing a Gaussian process prior on $f$ and then computing the
posterior for $f$ given $\mathcal{D}$ \cite{Rasmussen2006}. For this, let
$\mathbf{f}$ be the $n$-dimensional vector such that
$\mathbf{f}=(f(\mathbf{z}_1),\ldots,f(\mathbf{z}_n))^\text{T}$.  The prior for
$\mathbf{f}$ given $\mathcal{D}_\mathbf{Z}$ is \begin{equation}
p(\mathbf{f}|\mathcal{D}_\mathbf{Z}) =
\mathcal{N}(\mathbf{f}|\mathbf{m},\mathbf{K})\,,\label{eq:gpPrior}
\end{equation} where $\mathbf{m}$ is a $n$-dimensional mean vector and
$\mathbf{K}$ is an $n \times n$ covariance matrix generated by the covariance
function or kernel
\begin{align}
k_{ij} & \equiv \text{Cov}[f(\mathbf{z}_i), f(\mathbf{z}_j)] \nonumber \\
& \equiv  \sigma \exp \left\{ - (\mathbf{z}_i - \mathbf{z}_j)^\text{T}
\text{diag}(\boldsymbol{\lambda}) (\mathbf{z}_i - \mathbf{z}_j)\right\} + \sigma_0\,,\label{eq:covariance}
\end{align}
where $\boldsymbol{\lambda}$ is a vector of lengthscales and $\sigma$, $\sigma_0$ are
amplitude and noise parameters.  Then, the posterior distribution for
$\mathbf{f}$ given $\mathcal{D}_{U,V}$ and $\mathcal{D}_{\mathbf{Z}}$ is
\begin{equation}
p(\mathbf{f}|\mathcal{D}_{U,V},\mathcal{D}_\mathbf{Z}) =
\frac{p(\mathcal{D}_{U,V}|\mathbf{f})p(\mathbf{f}|\mathcal{D}_\mathbf{Z})}{p(\mathcal{D}_{U,V}|\mathcal{D}_{\mathbf{Z}})}\,,
\label{eq:posterior}
\end{equation}
where $p(\mathcal{D}_{U,V}|\mathbf{f}) = \prod_{i=1}^n c(u_i,v_i|\tau =
2\Phi(f_i) - 1)$, $p(\mathcal{D}_{U,V}|\mathcal{D}_{\mathbf{Z}})$ is a
normalization constant and $c(\cdot, \cdot | \tau)$ is the density of a
parametric bivariate copula specified in terms of Kendall's $\tau$.  Given a
particular value of $\mathbf{Z}$ such as $\mathbf{z}^\star$, we can make
predictions about the conditional distribution of $U$ and $V$ given
$\mathbf{z}^\star$ using
\begin{align}
p(u^\star,v^\star|\mathbf{z}^\star) = & \int c(u^\star,v^\star|\tau = 2\Phi(f^\star) - 1)\nonumber\\
& \hspace{-1cm} p(f^\star|\mathbf{f},\mathbf{z}^\star,\mathcal{D}_\mathbf{z})
p(\mathbf{f}|\mathcal{D}_{U,V},\mathcal{D}_\mathbf{Z})\,d\mathbf{f}\,df^\star\,,\label{eq:predictive}
\end{align}
$p(f^\star|\mathbf{f},\mathbf{z}^\star,\mathcal{D}_\mathbf{z})=
\mathcal{N}(f^\star|\mathbf{k}^\text{T}\mathbf{K}^{-1} \mathbf{f}, k - \mathbf{k}^\text{T} \mathbf{K}^{-1} \mathbf{k})$,
$\mathbf{k} = (\text{Cov}(f(\mathbf{z}^\star),f(\mathbf{z}_1)),\ldots,\text{Cov}(f(\mathbf{z}^\star),f(\mathbf{z}_n)))^\text{T}$ and
$k = \text{Cov}(f(\mathbf{z}^\star),f(\mathbf{z}^\star))$.
Unfortunately, (\ref{eq:posterior}) and (\ref{eq:predictive}) cannot be
computed analytically, so we decide to approximate them using Expectation Propagation (EP)
\cite{Minka2001}.  This method approximates each of the $n$ factors in
$p(\mathcal{D}_{U,V}|\mathbf{f})$ with an unnormalized Gaussian distribution
whose mean and variance parameters are updated iteratively by matching
sufficient statistics. See \citet{Rasmussen2006} for further details.  To refine
each of these univariate Gaussians, we have to compute three unidimensional
integrals using quadrature methods. For prediction at $\mathbf{z}^\star$, we
sample $f(\mathbf{z}^\star)$ from the Gaussian approximation found by EP and
then average over copula models with $\tau = 2 \Phi(f(\mathbf{z}^\star)) - 1$.
The resulting conditional copula model is semi-parametric: The dependence
between $U$ and $V$ given $\mathbf{Z}$ is parametric but the effect of
$\mathbf{Z}$ on the copula is non-parametric.

\subsection{Sparse GPs to Speed up Computations}

The total cost of EP is $O(n^3)$, since it is dominated by the computation of
the Cholesky decomposition of an $n\times n$ matrix.  To reduce this cost, we
use the FITC approximation for Gaussian Processes
\cite{Snelson2006,Naish-Guzman2007}. Under this approximation, the $n \times n$
covariance matrix $\mathbf{K}$ is approximated by $\mathbf{K}' =  \mathbf{Q} +
\text{diag}(\mathbf{K} - \mathbf{Q})$, where $\mathbf{Q} = \mathbf{K}_{nn_0}
\mathbf{K}_{n_0n_0}^{-1} \mathbf{K}_{nn_0}^\text{T}$, $\mathbf{K}_{n_0n_0}$ is
the $n_0 \times n_0$ covariance matrix generated by evaluating
(\ref{eq:covariance}) at all combinations of some $n_0 \ll n$ training points
or \emph{pseudo-inputs}, and $\mathbf{K}_{nn_0}$ is the $n \times n_0$ matrix
with the covariances between all possible combinations of original training
points and pseudo-inputs. These approximations allow us to run the EP method
with cost ${O}(nn_0^2)$. The kernel hyper-parameters $\boldsymbol{\lambda}$,
$\sigma$ and $\sigma_0$ and the pseudo-inputs are optimized by maximizing the
EP approximation of the model evidence \cite{Rasmussen2006}.

\begin{table}
  \caption{Some copula families, with their parameter domains and
  expressions as Kendall's $\tau$ correlations. For some copulas like Frank or
  Joe, numerical approximations are used to compute $\theta(\tau)$.}
  \begin{center}
  \begin{tabular}{|l|l|l|}
  \hline
  \textbf{Family} & \textbf{Parameter} & $\theta(\tau) = $ \\ \hline
  Gaussian               & $\theta \in [-1,1]$       & \multirow{2}{*}{$sin\left(\dfrac{\pi}{2} \tau\right)$} \\\cline{1-2}
  Student                & $\theta \in [-1,1]$       & \\ \hline
  Clayton                & $\theta \in (0, \infty) $ & $2\tau/(1-\tau)$ \\ \hline
  Gumbel                 & $\theta \in [1,\infty)$   & $1/(1-\tau)$ \\ \hline
  Frank                  & $\theta \in (0, \infty)$  & \multirow{2}{*}{No closed form} \\\cline{1-2}
  Joe                    & $\theta \in (1, \infty)$  & \\ \hline
  \end{tabular}
  \end{center}
  \label{table:thetatau}
\end{table}

Learning a vine with our method scales linearly with the number of samples $n$,
but quadratically with the number of pseudo-inputs $n_0$ and the number of variables
$d$: thus, its complexity is $O(d^2n_0^2n)$. By contrast,
learning a \emph{simplified} vine has complexity $O(d^2)$.

\subsection{Related Work}\label{sec:mll}

Acar, Genest and Neslehova \yrcite{Acar2012} first addressed the lack
of conditional dependencies in the parametric copulas that form a vine model.
They use a method similar in spirit to the one described above, and model
$\tau$ as a non-linear function of a single conditioning variable $Z$
\cite{Acar2011}.  However, their method cannot handle multivariate conditional
dependences and consequently, they only show results for trivariate vines.
They use the \emph{Maximum Local Likelihood} (MLL) method to infer a non-linear
relationship between $\tau$ and $Z$. Acar et al.
approximate linearly $f$ at any point $z$ where $f$ needs to be evaluated.
Then, they adjust the coefficients of the resulting linear form using the
available observations in the neighborhood of $z$.  In particular, given a
sample $\mathcal{D} = \{ \mathcal{D}_{U,V} = \{(u_i, v_i)\}_{i=1}^n,
\mathcal{D}_Z = \{z_i\}_{i=1}^n\}$, they estimate $f(z)$ by solving the
optimization problem
\begin{align}
(b_{z,0}^\star, b_{z,1}^\star) &= \operatorname*{arg\,max}_{(b_0, b_1) } \left\{ \sum_{i=0}^N k_h(z-z_i)  \right.\nonumber\\
& \left. \log c(u_i,v_i| \tau = 2 \Phi(b_0 + b_1(z-z_i)) - 1)\right\},\label{eq:elif}
\end{align}
where the neighborhood of $z$ is determined by the Epanechnikov kernel $k_h(x)
= \frac{3}{4h}\max (0, 1-(\frac{x}{h})^2)$ with bandwidth $h$. An estimate of
$f(z)$ is then obtained as the intercept of the linear approximation at $z$,
that is, $f(z) \approx b_{z,0}^\star$.  Acar et al. adjust $h$ by running a
leave-one-out cross validation search on the training data.  Some disadvantages
of the MLL method are: (i) it can only condition on a single scalar variable,
(ii) we have to solve the optimization problem (\ref{eq:elif}) for each
prediction that we want to make and more importantly (iii) since it is a
local-based method (similarly as nearest neighbours) it can lead to poor
predictive performance when the available data is sparsely distributed.

\section{Experiments}\label{sec:experiments}

\begin{figure*}
  \includegraphics[width=\textwidth]{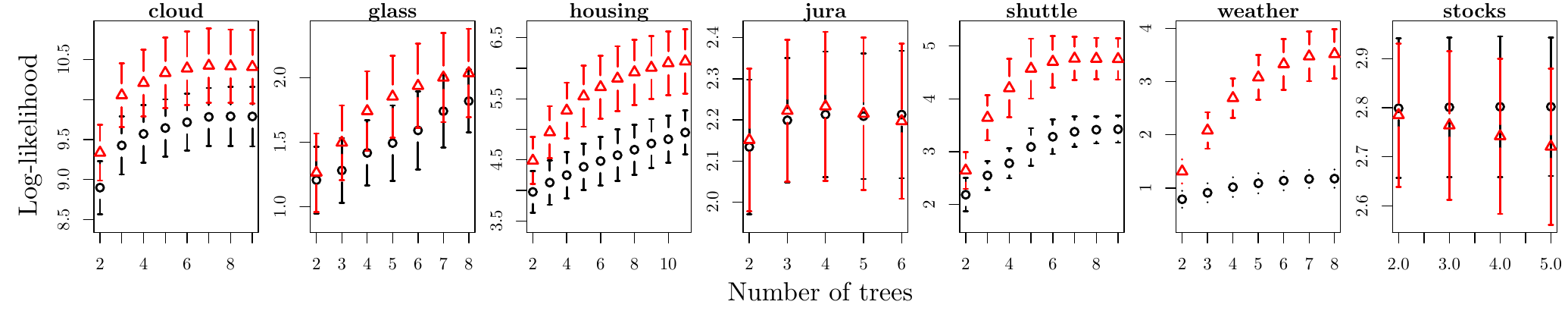}
  \vskip -0.3 cm
  \caption{Average test log-likelihood and standard deviations for each dataset
  as the number of trees forming the vines increases, for GPVINE (red
  triangles) and SVINE (black dots) (higher is better).}
  \label{fig:bars}
\end{figure*}

\begin{table}
  \begin{center}
  \caption{Average test log-likelihood and standard deviations for SVINE and
  GPVINE on real-world datasets (higher is better). Asterisks denote results that are not
  statistically significant to a paired Wilcoxon test with $\text{p--value} =
  10^{-3}$.\\}\label{table:realexps}
  \resizebox{\linewidth}{!}{
  \begin{tabular}{lccc}
  \hline
  \textbf{Data} & \textbf{Trees} & \textbf{SGVINE} & \textbf{GPVINE}\\
  \hline
    \multirow{9}{*}{Cloud}   & 1  & $\f{7.860 \pm 0.346}$ & $\f{ 7.860 \pm 0.346}$\\
                             & 2  & $   8.899 \pm 0.334 $ & $\f{ 9.335 \pm 0.348}$\\
                             & 3  & $   9.426 \pm 0.363 $ & $\f{10.053 \pm 0.397}$\\
                             & 4  & $   9.570 \pm 0.361 $ & $\f{10.207 \pm 0.415}$\\
                             & 5  & $   9.644 \pm 0.357 $ & $\f{10.332 \pm 0.440}$\\
                             & 6  & $   9.716 \pm 0.354 $ & $\f{10.389 \pm 0.459}$\\
                             & 7  & $   9.783 \pm 0.361 $ & $\f{10.423 \pm 0.463}$\\
                             & 8  & $   9.790 \pm 0.371 $ & $\f{10.416 \pm 0.459}$\\
                             & 9  & $   9.788 \pm 0.373 $ & $\f{10.408 \pm 0.460}$\\\hline
    \multirow{8}{*}{Glass}   & 1  & $\f{0.827 \pm 0.150}$ & $\f{ 0.827 \pm 0.150}$\\
                             & 2  & $  {1.206 \pm 0.259}$ & $\f{ 1.264 \pm 0.303}$\\
                             & 3  & $  {1.281 \pm 0.251}$ & $\f{ 1.496 \pm 0.289}$\\
                             & 4  & $  {1.417 \pm 0.251}$ & $\f{ 1.740 \pm 0.308}$\\
                             & 5  & $  {1.493 \pm 0.291}$ & $\f{ 1.853 \pm 0.318}$\\
                             & 6  & $  {1.591 \pm 0.301}$ & $\f{ 1.936 \pm 0.325}$\\
                             & 7  & $  {1.740 \pm 0.282}$ & $\f{ 2.000 \pm 0.345}$\\
                             & 8  & $  {1.818 \pm 0.243}$ & $\f{ 2.034 \pm 0.343}$\\\hline
    \multirow{6}{*}{Jura}    & 1  & $\f{1.887 \pm 0.153}$ & $\f{ 1.887 \pm 0.153}$\\
                             & 2  & $  {2.134 \pm 0.164}$ & $\f{ 2.151 \pm 0.173}$\\
                             & 3  & $  {2.199 \pm 0.151}$ & $\f{ 2.222 \pm 0.173}$\\
                             & 4*  & $  {2.213 \pm 0.153}$ & $\f{ 2.233 \pm 0.181}$\\
                             & 5*  & $  {2.209 \pm 0.153}$ & $\f{ 2.215 \pm 0.185}$\\
                             & 6*  & $\f{2.213 \pm 0.155}$ & $  { 2.197 \pm 0.189}$\\\hline
    \multirow{9}{*}{Shuttle} & 1  & $\f{1.487 \pm 0.256}$ & $\f{ 1.487 \pm 0.256}$\\
                             & 2  & $  {2.188 \pm 0.314}$ & $\f{ 2.646 \pm 0.349}$\\
                             & 3  & $  {2.552 \pm 0.273}$ & $\f{ 3.645 \pm 0.427}$\\
                             & 4  & $  {2.782 \pm 0.284}$ & $\f{ 4.204 \pm 0.551}$\\
                             & 5  & $  {3.092 \pm 0.353}$ & $\f{ 4.572 \pm 0.567}$\\
                             & 6  & $  {3.284 \pm 0.325}$ & $\f{ 4.703 \pm 0.492}$\\
                             & 7  & $  {3.378 \pm 0.288}$ & $\f{ 4.763 \pm 0.408}$\\
                             & 8  & $  {3.417 \pm 0.257}$ & $\f{ 4.761 \pm 0.393}$\\
                             & 9  & $  {3.426 \pm 0.252}$ & $\f{ 4.755 \pm 0.389}$\\\hline
    \multirow{8}{*}{Weather} & 1  & $\f{0.684 \pm 0.128}$ & $\f{ 0.684 \pm 0.128}$\\
                             & 2  & $  {0.789 \pm 0.159}$ & $\f{ 1.312 \pm 0.227}$\\
                             & 3  & $  {0.911 \pm 0.178}$ & $\f{ 2.081 \pm 0.341}$\\
                             & 4  & $  {1.017 \pm 0.184}$ & $\f{ 2.689 \pm 0.368}$\\
                             & 5  & $  {1.089 \pm 0.188}$ & $\f{ 3.078 \pm 0.423}$\\
                             & 6  & $  {1.138 \pm 0.181}$ & $\f{ 3.326 \pm 0.477}$\\
                             & 7  & $  {1.170 \pm 0.169}$ & $\f{ 3.473 \pm 0.467}$\\
                             & 8  & $  {1.177 \pm 0.170}$ & $\f{ 3.517 \pm 0.465}$\\\hline
    \multirow{5}{*}{Stocks}  & 1  & $\f{2.776 \pm 0.142}$ & $  { 2.776 \pm 0.142}$\\
                             & 2*  & $\f{2.799 \pm 0.142}$ & $  { 2.785 \pm 0.146}$\\
                             & 3  & $\f{2.801 \pm 0.142}$ & $  { 2.764 \pm 0.151}$\\
                             & 4  & $\f{2.802 \pm 0.143}$ & $  { 2.742 \pm 0.158}$\\
                             & 5  & $\f{2.802 \pm 0.141}$ & $  { 2.721 \pm 0.159}$\\\hline
    \multirow{6}{*}{Housing} & 1  & $\f{3.409 \pm 0.354}$ & $\f{ 3.409 \pm 0.354}$\\
                             & 3  & $  {4.128 \pm 0.363}$ & $\f{ 4.953 \pm 0.425}$\\
                             & 5  & $  {4.386 \pm 0.380}$ & $\f{ 5.541 \pm 0.498}$\\
                             & 7  & $  {4.576 \pm 0.422}$ & $\f{ 5.831 \pm 0.529}$\\
                             & 9  & $  {4.768 \pm 0.399}$ & $\f{ 6.009 \pm 0.516}$\\
                             & 11 & $  {4.949 \pm 0.362}$ & $\f{ 6.113 \pm 0.525}$\\\hline
  \end{tabular}}
  \end{center}
\end{table}

We evaluate the performance of the proposed method for the
estimation of vine copula densities with full conditional dependencies.  Because GPs are
an important part in this method we call it GPVINE.  We compare with two
benchmark methods: (i) a vine model based on the simplifying assumption
(SVINE), which ignores any conditional dependencies in the bivariate copulas,
and (ii) a vine model based on the MLL method of \citet{Acar2012} (MLLVINE).
MLLVINE can only handle conditional dependencies with respect on a single
scalar variable.  Therefore, we can only evaluate its performance in the
construction of vine models with two trees, since additional levels would
require to account for multivariate conditional dependencies.

In all the experiments, we use 20 pseudo-inputs in the generalized FITC
approximation. The Gaussian processes use a kernel function given by
(\ref{eq:covariance}), whose hyper-parameters and pseudo-inputs are tuned by
maximizing the EP estimate of the marginal likelihood.  The mean of the GP
prior (\ref{eq:gpPrior}) is chosen to be constant and equal to
$\Phi^{-1}((\hat{\tau}_{MLE} + 1) / 2)$, where $\hat{\tau}_{MLE}$ is the
maximum likelihood estimate of $\tau$ for an unconditional Gaussian copula
given the training data.  In MLLVINE, the bandwidth of the Epanechnikov kernel
is selected by running a leave-one-out cross validation search using a
30-dimensional log-spaced grid ranging from 0.05 to 10.  To simplify the
experiments, we focus on regular vines generated using bivariate Gaussian
copulas (see Appendix \ref{appendix:gaussiancop}) as building blocks.  The
extension of the proposed approach to select among different parametric
families of bivariate copulas is straightforward: the best family for a given
pair of variables can be selected by Bayesian model selection, using the
evidence approximation given by EP.  All the data are preprocessed to have
uniform marginal distributions: this is done by mapping each marginal
observation to its empirical cumulative probability.

\subsection{Synthetic Data}\label{exp:syn}

We first perform a series of experiments with synthetic three-dimensional data.
In particular, we sample the scalar variables $X$, $Y$ and $Z$ according to the
following generative process.  First, $Z$ is sampled uniformly from the
interval $[-6,6]$ and second, $X$ and $Y$ are sampled given $Z$ from a
bivariate Gaussian distribution with zero mean and covariance matrix given by
$\text{Var}(X) = \text{Var}(Y) = 1$ and $\text{Cov}(X,Y|Z) = 3 / 4
\sin(Z)$.  We sample a total of 1000 data points and then choose 50 subsamples
of size 100 to infer a vine model for the data using SVINE, MLLVINE and GPVINE.
Average test log-likelihoods on the remaining data points are shown in Table
\ref{table:little_exps}. GPVINE obtains the best results.

\begin{figure}
  \begin{center}
    \vskip -0.7 cm
    \includegraphics[width=\linewidth]{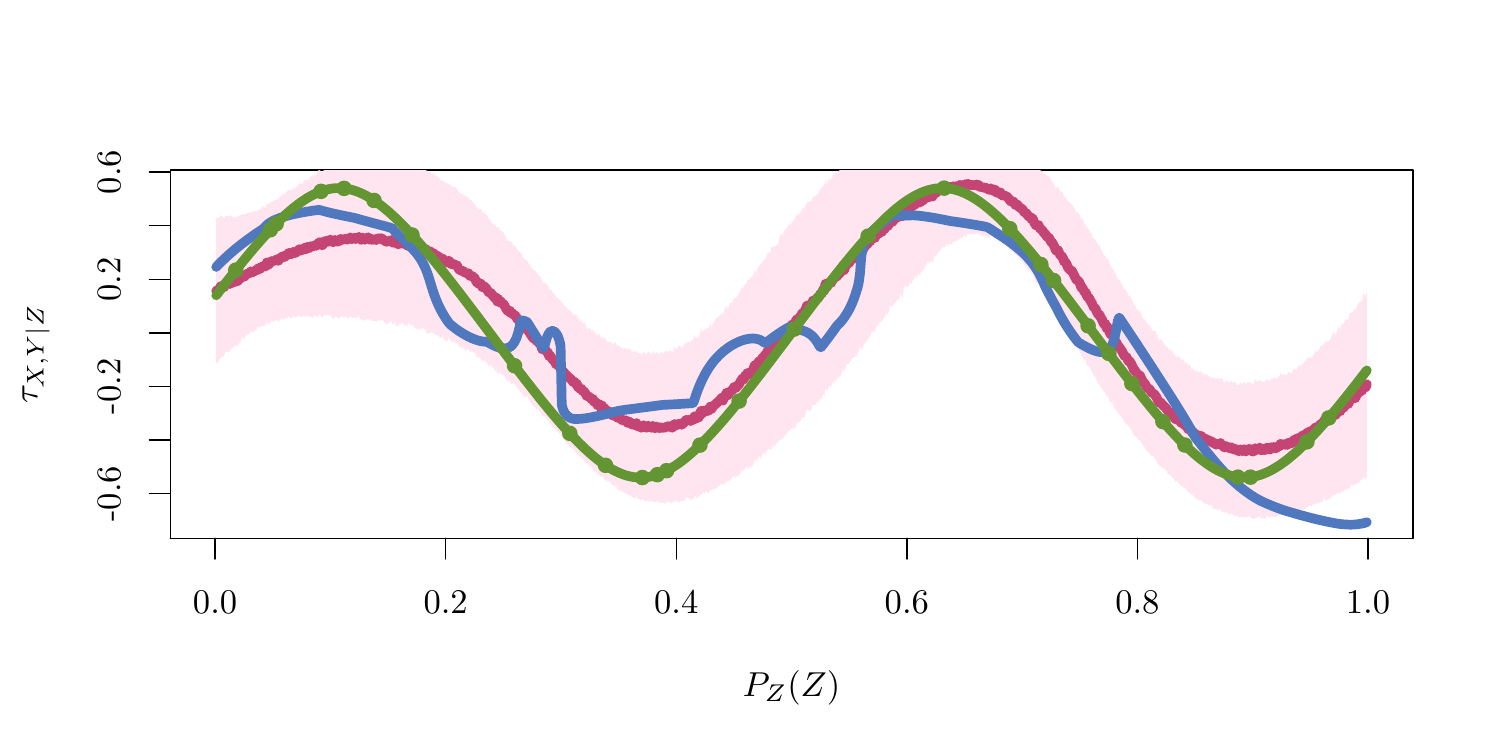}
    \vskip -0.3 cm
    \caption{In green, the true function $g$ that maps $u_3$ to $\tau$.  The
    approximations learned by GPVINE and MLLVINE are shown in red and blue,
    respectively.  For GPVINE, the uncertainty of the prediction ($\pm$1
    standard deviation) is drawn as a red shaded area.  Small dots are used to
    mark the available observations of $u_3$ for training.}
    \label{fig:syn}
  \end{center}
\end{figure}

Figure \ref{fig:syn} displays the true value of the function $g$ that maps
$u_3$ to $\tau$ in the conditional copula $c(P(u_1|u_3),P(u_1|u_3)|u_3,\tau)$, where
$u_1$, $u_3$ and $u_3$ are the empirical cumulative probability levels of the
samples generated for $X$, $Y$ and $Z$, respectively. We also show the
approximations of $g$ generated by GPVINE and MLLVINE.  In this case, GPVINE is
much better than MLLVINE at approximating the true $g$.

\subsection{Real-world Datasets}

We evaluate the performance of SVINE, MLLVINE and GPVINE on several real-world
datasets. For each dataset, we generate 50 random partitions of the data into
training and test sets, each containing half of the available data.  The
different methods are run on each training set and their log-likelihood is then
evaluated on the corresponding test set (higher is better). The analyzed
datasets are described in Section \ref{sec:datasets}.  Table \ref{table:realexps} and Figure
\ref{fig:bars} show the test log-likelihood for SVINE and GPVINE, when using up
to $1,\ldots,(d-1)$ trees in the vine, where $d$ is the number of variables in
the data.  In general, taking into account possible dependencies in the
conditional bivariate copulas leads to superior predictive performance.  Also,
we often find that the gains obtained get larger as we increase the number of
trees in the vines.  However, in a few of the datasets the simplifying
assumption seems valid (Stocks and Jura datasets).  Table
\ref{table:little_exps} shows results for all methods (including MLLVINE) when
only two trees are used in the vines. In these experiments, MLLVINE is most of
the times outperformed by GPVINE. To better measure the percent 
improvement experienced when using GPVINE, one can subtract the achieved
likelihood when using only the first tree of the vine from the all results. We
also show how GPVINE can be used to discover scientifically interesting
features through learning spatially varying correlations (Figure
\ref{fig:bcn_tau}).

\subsubsection{Description of the Datasets}\label{sec:datasets}

\paragraph{Mineral Concentrations}
The \emph{jura} dataset contains the concentration measurements of 7 chemical
elements (Cd, Co, Cr, Cu, Ni, Pb, Cn) in 359 locations of the Swiss Jura
Mountains \citep{jura}. The \emph{uranium} dataset contains log-concentrations
of 7 chemical elements (U, Li, Co, K, Cs, Sc, Ti) in a total of 655 water
samples collected near Grand Junction, CO \citep{uranium}.  \citet{Acar2012}
use the measurements for $Co$, $Ti$ and $Sc$ to evaluate the performance of
MLLVINE. We replicated this task for the three analyzed methods (Table
\ref{table:little_exps}).

\begin{table}
  \caption{Average test log-likelihood and standard deviations for all methods and datasets
  when limited to 2 trees in the vine (higher is better).\\}\label{table:little_exps}
  \centering
  \resizebox{\linewidth}{!}{
  \begin{tabular}{lccc}
  \hline
  \textbf{Data} & \textbf{SVINE} & \textbf{MLLVINE} & \textbf{GPVINE}\\ \hline
  Synthetic   & $-0.005 \pm 0.012$ & $0.101 \pm 0.162 $ & $  \f{ 0.298 \pm 0.031}$ \\
  Uranium     & $ 0.006 \pm 0.006$ & $0.016 \pm 0.026$   & $ \f{ 0.022 \pm 0.012}$ \\
  Cloud       & $ 8.899 \pm 0.334$ & $9.013 \pm 0.600$   & $ \f{ 9.335 \pm 0.348}$ \\
  Glass       & $ 1.206 \pm 0.259$ & $0.460 \pm 1.996$   & $ \f{ 1.264 \pm 0.303}$ \\
  Housing     & $ 3.975 \pm 0.342$ & $4.246 \pm 0.480$   & $ \f{ 4.487 \pm 0.386}$ \\
  Jura        & $ 2.134 \pm 0.164$ & $2.125 \pm 0.177$   & $ \f{ 2.151 \pm 0.173}$ \\
  Shuttle     & $ 2.552 \pm 0.273$ & $2.256 \pm 0.612$   & $ \f{ 3.645 \pm 0.427}$ \\
  Weather     & $ 0.789 \pm 0.159$ & $0.771 \pm 0.890$   & $ \f{ 1.312 \pm 0.227}$ \\
  Stocks      & $ \f{2.802 \pm 0.141}$ & $2.739 \pm 0.155$   & $ { 2.785 \pm 0.146}$ \\
  \hline
  \end{tabular}
  }
\end{table}

\begin{figure}[b!]
  \begin{center}
    \vskip -0.8 cm
    \includegraphics[width=\linewidth]{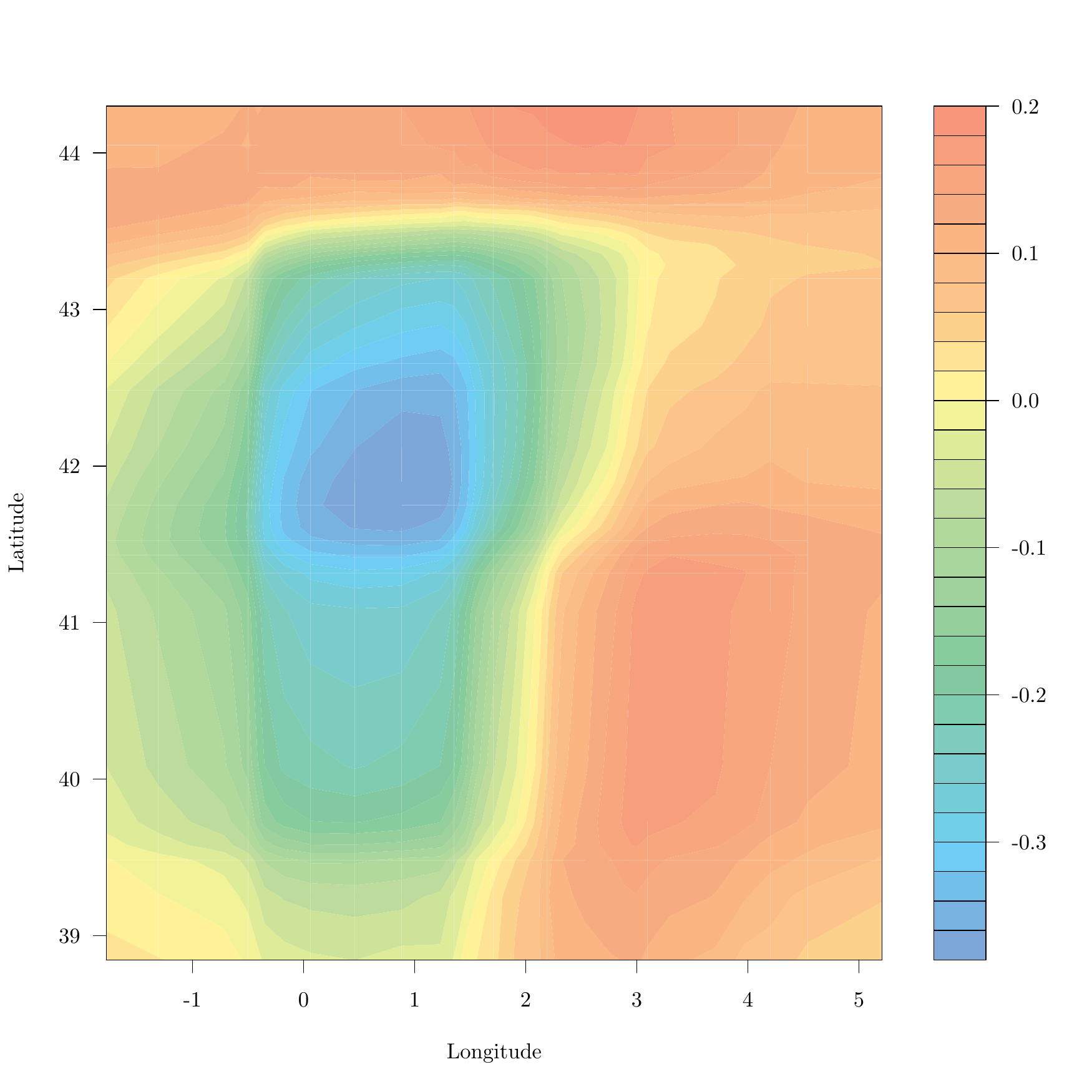}
    \vskip -0.3 cm
    \caption{Kendall's $\tau$ correlation between \emph{atmospheric
    pressure} and \emph{cloud percentage cover} (color scale) when conditioned
    to \emph{longitude} and \emph{latitude}.  The blue region
    in the plot corresponds to the Pyrenees mountains.}
    \label{fig:bcn_tau}
  \end{center}
\end{figure}


\paragraph{Barcelona Weather}
\emph{OpenWeatherMap} \cite{open_weather} provides access to meteorological
stations around the world. We downloaded data for the 300 weather stations
nearest to Barcelona, Spain (41.3857N, 2.1699E) on 11/19/2012 at 8pm
(\emph{weather} dataset). Each station returns values for \emph{longitude},
\emph{latitude}, \emph{distance to Barcelona}, \emph{temperature},
\emph{atmospheric pressure}, \emph{humidity}, \emph{wind speed}, \emph{wind
direction} and \emph{cloud cover percentage}.  Figure \ref{fig:bcn_tau} shows
how the posterior mean of $\tau$ for the copula linking the variables
\emph{atmospheric pressure} and \emph{cloud cover percentage} varies when
conditioned on \emph{latitude} and \emph{longitude}.

\paragraph{World Stock Indices}
We apply the probability integral transform to the residuals of an
ARMA(1,1)-GARCH(1,1) model with Student $t$ innovations.  The residuals are
obtained after fitting this model to the daily log-returns of the major world stock
indices in 2009 and 2010 (\emph{stocks} dataset, 396 points in total)
\cite{rvines}.  The considered indices are the US American \emph{S\&P 500}, the
Japanese \emph{Nikkei 225}, the Chinese \emph{SSE Composite Index}, the German
\emph{DAX}, the French \emph{CAC 40} and the British \emph{FTSE 100 Index}.

\paragraph{UCI Datasets}
We also include experimental results for the \emph{Glass}, \emph{Housing},
\emph{Cloud} and \emph{Shuttle} datasets from the UCI Dataset Repository
\cite{uci}.


%
%

\section{Conclusion}\label{sec:conclusions}
Vine copulas are increasingly popular models for multivariate data. They 
specify a factorization of any high-dimensional copula density into a product of
conditional bivariate copulas. However, some of the conditional dependencies in
these bivariate copulas are usually ignored when constructing the vine. This
can produce overly simplistic estimates when dealing with real-world data. To
avoid this, we presented a method for the estimation of fully conditional
vines using Gaussian processes (GPVINE). A series of experiments with synthetic
and real-world data show that, often, GPVINE obtains better
predictive performance than a baseline method that ignores conditional
dependencies. Additionally, GPVINE performs favorably
with respect to state-of-the-art alternatives based on maximum local-likelihood
methods (MLLVINE).

\section*{Acknowledgements}

DLP and JMLH contributed equally to this work. We want to thank Philipp Hennig
and Andrew Gordon Wilson for their helpful feedback. DLP was funded by
Fundaci\'on Caja Madrid and the PASCAL2 Network of Excellence. JMHL was funded
by Infosys Labs, Infosys Limited. 

\appendix

\section{The Bivariate Gaussian Copula}\label{appendix:gaussiancop}
The bivariate Gaussian copula with correlation parameter $\theta$ represents
the dependence structure found in a bivariate Gaussian distribution of two
random variables with correlation $\theta$. The Gaussian copula has cdf
\begin{equation}
  C(u,v| \theta) = \Phi_2(\Phi^{-1}(u),\Phi^{-1}(v))| \theta),
  \end{equation}
where $\Phi_2(\cdot,\cdot|\theta)$ is the cdf of a bivariate Gaussian with
marginal variances equal to one and correlation $\theta$, and $\Phi^{-1}$ is the
quantile function of the standard Gaussian distribution.  The corresponding pdf
is
\begin{equation}
  c(u,v| \theta) = \frac{\phi_2(\Phi^{-1}(u),
  \Phi^{-1}(v)|\theta)}{\phi(\Phi^{-1}(u))\phi(\Phi^{-1}(v))},
\end{equation}
where $\phi_2$ is the derivative (pdf) of $\Phi_2$. The conditional cdfs are
given by
\begin{align}
  \frac{\partial C(u,v| \theta)}{\partial u}  &=
  \Phi\left(\frac{\Phi^{-1}(v) - \theta \, \Phi^{-1}(u)}{\sqrt{1-\theta^2}}\right)\,, \\
  \frac{\partial C(u,v| \theta)}{\partial v}  &=
  \Phi\left(\frac{\Phi^{-1}(u) - \theta \, \Phi^{-1}(v)}{\sqrt{1-\theta^2}}\right),
\end{align}
where $\Phi$ is the standard Gaussian cdf.

\clearpage
\newpage
\bibliography{Bibliography}
\bibliographystyle{icml2013}

\end{document}